\documentclass[11pt]{article}
\usepackage[dvips]{graphics}
\def\oneandahalfspace{\baselineskip=\normalbaselineskip
  \multiply\baselineskip by 3 \divide\baselineskip by 2}

\newcommand{\Z}{\hbox{{\rm /}\kern-.4em\hbox{\rm Z}}}

\begin{document}
\oneandahalfspace

\title{Monte Carlo simulation of a two-field effective Hamiltonian of
 complete wetting}

\author{S. Flesia\\
Department of Mathematics, Imperial College,\vspace*{1ex}\\ 
180 Queens Gate, London SW7 2BZ, U.K\vspace*{1ex}\\
}

\date{\today} 

\maketitle

\begin{abstract}
\oneandahalfspace

Recent work on the complete wetting transition for three dimensional systems
with short-ranged forces has 
emphasized the role played by the coupling of order-parameter 
fluctuations near the wall and  depinning interface.
It has been proposed that an effective two-field
Hamiltonian, which predicts a renormalisation of the wetting parameter,
could explain the controversy between RG analysis of the capillary-wave 
model and Monte Carlo simulations on 
the Ising model. In this letter results of extensive Monte Carlo simulations 
of the two-field model are presented. The results are in agreement with
prediction of a renormalized wetting parameter $\omega $.
 
\end{abstract} 




\newpage

There are a number of long standing controversies in the study of phase 
equilibria at fluid interfaces, related to wetting transitions for systems 
with short-ranged forces at the marginal dimension $d=3$ (for a recent 
review see for example \cite{parry}).
The standard model used to describe fluctuation effects at wetting transitions
is the effective interfacial Hamiltonian (also called the capillary wave (CW) 
model)
\begin{equation}
\label{capillary}
 H[\ell({\bf y})]=\int d{\bf y} \Bigl[\frac{\Sigma_{\alpha \beta } (T)}{2} 
(\nabla \ell)^2 + W(\ell)\Bigr]
\end{equation}
where $\ell({\bf y}) >0 $ is a collective coordinate which represents the 
distance of the fluctuating fluid interface (separating bulk phases 
$\alpha $ and $\beta $, say) from a wall situated in the $\ell =0$ plane.
Here $\Sigma_{\alpha \beta }$ is the 
interfacial stiffness coefficient and $W(\ell)$ is the effective binding 
potential which is usually specified as
\begin{equation}
\label{potential}
W(\ell)=A e^{-\kappa \ell} + B e^{-2\kappa \ell} + \bar{h}\ell
\end{equation}
where $\kappa \equiv 1/\xi_b $ is the inverse bulk correlation length 
and $\bar{h}$ is proportional to a bulk ordering field.
At a critical or complete wetting transition the mean interface displacement
$\langle \ell \rangle $ (and other lengthscales) diverges as the external field
approach critical values.
For critical wetting, B is positive and the transition occurs at $\bar{h}=A=0$,
corresponding to $T\rightarrow T_W $, the wetting temperature,
while for complete wetting $A > 0$, i.e. $T > T_W $ and the transition 
occurs for $\bar{h} \rightarrow 0 $ for $T_C > T > T_W$.\\
One well known controversy concerns the critical wetting transition.
For this case, RG calculations based on (\ref{capillary}) (see for example 
\cite {brezin}), predict strong 
non-universality for the critical exponents and amplitudes depending on the 
value of the wetting parameter $\omega (T)=k_B T \kappa^2/4\pi 
\Sigma_{\alpha \beta } $ at the transition temperature $T_W $. 
However, extensive Monte Carlo simulations of wetting in a three 
dimensional Ising model with nearest-neighbour interactions by 
Binder, Landau and Kroll \cite{BLK86}, show only small deviations from 
mean-field theory consistent with $\omega \approx 0 $. 
This contrasts sharply with the expected value $\omega \approx 0.8 $ 
(at the temperature where the simulations were performed). 
Prompted by the discrepancy Gompper and Kroll [GK] \cite{GK87} performed very 
accurate Monte Carlo simulations on a lattice version of (\ref{capillary}) 
and found very good agreement with the renormalisation group treatment, 
leaving the controversy open.\\
More recently attention has focused on the complete wetting transition. 
For this case CW theory predicts that, while the critical exponents retain 
their mean-field values in the limit of $\bar{h} \rightarrow 0 $, 
critical amplitudes are renormalised. For example, the mean distance 
$\langle \ell \rangle $ diverges at leading order as
\begin{equation}
\label{meanl}
\kappa \langle \ell \rangle \approx \theta \ln (1/\bar{h})
\end{equation}
where $\theta $ depends on $\omega $. In particular
for $\omega < 2 $, CW theory predicts $\theta^{CW}=1+\frac{\omega }{2}$.
A new controversy appeared with the Ising model simulations
of Binder, Landau and Ferrenberg \cite{BLF95}, 
(at $\omega \sim 0.8 $) consistent with $\theta \sim 1.6 -1.8$, while 
the CW model predicts only $\theta^{CW} \approx 1.4 $.

These results serve to emphasize that wetting at the marginal dimension 
is a rather special case and that the coarse-grained Hamiltonian 
(\ref{capillary}) may not capture all the essential physics. 
Note that no such problems arise in two
dimensions where the CW model yields predictions for universal critical 
exponents, amplitudes and scaling functions which are in excellent agreement
with the exact Ising model results \cite{abraham}.
Motivated by this, several authors have undertaken a careful 
examination of the foundations of the RG theory and of the validity of the 
interfacial Hamiltonian. In particular Fisher and Jin [FJ] \cite{fisherjin}
initiated a method of systematically deriving interfacial Hamiltonian
from an underlying  Landau-Ginzburg-Wilson [LGW] model.
For critical wetting, they suggested that (\ref{capillary}) should  
be modified by allowing a position dependent stiffness coefficient and 
as a result argue that the transition may in fact be very weakly first-order. 
More recently, Parry and Boulter [PB] \cite {PB} argued that this is still not
a sufficient modification of the CW model, since it does not account for
fluctuations of the order parameter near the wall and depinning 
interface.
Following the methods by [FJ]\cite{fisherjin}, they introduced 
a two-field Hamiltonian for complete wetting
\begin{equation}
\label{twoham}
 H[\ell_1,\ell_2]=\int d{\bf y}\{\frac{1}{2}\Sigma_{\mu \nu}(\ell_1,\ell_2) 
\nabla \ell_{\mu}\cdot \nabla \ell_{\nu} + U(\ell_1)+ 
W(\ell_2-\ell_1)\}
\end{equation}
which is a functional of two collective coordinates $\ell_1({\bf y})$ and 
$\ell_2({\bf y})$ (representing the locations of generalised surfaces at 
the wall and the $\alpha \beta $ interface respectively) and
${\bf \Sigma }$ is a position dependent stiffness matrix \cite {PB}. 
The potential 
$U(\ell_1)$ binds the lower surface to the wall and since the fluctuations 
of $\ell_1$ are small can be approximated by 
$U(\ell_1)=\frac{1}{2}v_0 \ell_1^2$. 
Here $v_0 = \Sigma_{11} /\xi_{w\beta}^2 $, where $\xi_{w\beta}$ is the
the finite correlation length associated with the intrinsic fluctuations of 
the order parameter near the wall. The fluctuations of $\ell_1({\bf y})$ 
are not included in the CW and FJ models.
The term $W(\ell_2-\ell_1)$ is similar to the binding potential in 
(\ref{potential}).
In the limit of complete wetting, $\ell_2$ unbinds from the wall
while $\ell_1$ remains bound. In their RG analysis, [PB] show show that 
the coupling to a weakly fluctuating 
field results in an effective value of the wetting parameter.
In the limit $v_0 \rightarrow \infty$ or $\xi_{w\beta} \rightarrow 0$ 
the fluctuations of the lower surface are completely suppressed and the 
CW/FJ result $\theta^{CW}=1+\frac{\omega }{2}$ is recovered. 
In the other limit, $v_0 \rightarrow 0$ or 
$\xi_{w\beta} \rightarrow \infty $, the Hamiltonian 
(\ref{twoham}) can be written in terms of center of mass coordinates which 
gives an effective stiffness coefficient $\Sigma^{-1}=\Sigma_{22}^{-1} 
+ \Sigma_{11}^{-1}$ and results in a capillary parameter 
$\omega_{eff}=\omega_2 + \omega_1 $.
For intermediate values of $v_0$, linear \cite{physicaAII}
and non-linear RG analysis \cite{BP96} predict that the effective 
capillary parameter is   
\begin{equation}
\label{omega}
\omega_{eff} =\frac{k_B T\kappa^2}{4\pi \Sigma_{22 }} 
+\frac{ k_B T\kappa^2}{4\pi (\Sigma_{11}+v_0/\Lambda^2 )}
\end{equation}
Consequently, the critical amplitude $\theta $ is given by 
$\theta=\theta^{CW} + \Delta \theta $ with 
\begin{equation}
\label{theta}
\Delta \theta=\frac{1}{2} \Bigl(\frac{ k_B T\kappa^2}{4\pi (\Sigma_{11}+ 
v_0/\Lambda^2 )}\Bigr).
\end{equation}
where $\Lambda $ is a momentum cutoff.
They estimated \cite{physicaAII}
$\Delta \theta \approx 0.3 \pm 0.1 $, 
which gives the prediction  $\theta \sim 1.7 \pm 0.1$, in reasonable agreement 
with the Ising simulations in the complete wetting regime. 
For the subsequent development in theory, it seems important that [PB] 
theory is tested by independent methods.
The crucial prediction is that the capillary 
parameter is renormalised due to the coupling to a field 
$\ell_1({\bf y})$ which only weakly fluctuates. 
To this end, in this letter, the Monte Carlo studies of critical wetting by 
[GK] \cite{GK87} have been generalised to
the two-field Hamiltonian in the complete wetting regime. 

The simulations were performed on a lattice version of (\ref{twoham}).
The two surfaces $\ell_1$ and $\ell_2$ are modelled by two parallel 
 $L \times L $ square lattices with periodic boundary conditions. 
The variables $x(i,j)$ and $y(i,j)$ are the distances of the site 
$(i,j)$ of the lower ($\ell_1$) and upper ($\ell_2$) surface respectively, 
from the wall (located at zero). 
The uncoupled gradient terms are approximated by
\begin{equation}
\label{grad}
\frac{1}{16 \pi \omega_1 } \sum_{(i,j),\delta} (x(i,j)-x_{\delta}
(i,j))^2, \ \ \ \ \ \  \frac{1}{16 \pi \omega_2 } \sum_{(i,j),\delta} 
(y(i,j)-y_{\delta}(i,j))^2 
\end{equation}
where $\delta $ indicate the four nearest-neighbours of the site $(i,j)$ and 
the sum is over all sites $(i,j)$ for $i,j=1,...,L$, and $x$ and $y$ 
are measured in units of the bulk correlation length $\xi_b $. 
To test the program, the potential (\ref{potential}) is first chosen as 
$A=0, B=1$ (critical wetting), $\Sigma_{11} $ is set very large to suppress 
the fluctuations of the lower surface $\ell_1$, the cross-gradient coupling 
term $\Sigma_{12}$ is kept to zero and $\omega_2 $ takes the values given 
by [GK] \cite{GK87}. The results are in very good agreement with their 
founding. For complete wetting, the potential is chosen
as $A=1$ and $B=0$ and the cross-gradient coupling term $\Sigma_{12}$ kept 
to zero \cite{physicaAII}.
The lattice size $L$ is varied between $10$ and $40$ 
to study finite-size effects. The SOS approximation is adopted and $x(i,j)$
and $y(i,j)$ are treated as continuous variables to avoid 
problems related to roughening. For simplicity we ignore the 
exponentially small position-dependence of 
the stiffness matrix elements since they play no role in determining
the increment $\Delta \theta $ \cite{physicaAII}.
The hard wall condition ($y \ge x$) is used here, although 
the prediction (\ref{omega}) is also valid in a soft wall approximation 
\cite{physicaAII,BP96} (see for example \cite{GK87} for a discussion 
regarding the CW model).
The wetting parameters are chosen as $\omega_1 = 1$ and  $\omega_2 = 0.8$, 
and so the maximum increment to $\theta $ is $\Delta \theta ^{max}=0.5$ 
(corresponding to $v_0=0$).\\
A Metropolis algorithm has been used on a Dec Alpha 3600.
At each Monte Carlo sweep, one of the two 
surfaces is chosen at random $2L^2$ times,
and then a randomly picked site $(i,j)$ is updated by a random number 
in the interval $[-z_m,+z_m]$; $z_m$ is set so that approximately 
50\% of the updates are accepted (but other values have also been 
used to test the convergence rate).
The first $10^4$ sweeps are discarded to thermalise the system 
and $10^6$ sweeps are 
used to calculate the averages. The block average method is used to obtain 
statistical errors and the integrated autocorrelation time $\tau_{int} $ is
calculated as in Madras and Sokal \cite{alan}.
For example for the parameter $v_0=100$ and field $h=0.1$, $\tau_{int} =15$ 
and for $h=0.01$, $\tau_{int} =113$, while for $v_0=0.05$ and field $h=0.1$, 
$\tau_{int} =25$ and for $h=0.01$, $\tau_{int} =159$. 
For a given value of the parameter $v_0$ the average distance
$\langle y \rangle$ of the upper surface from the wall, is calculated 
for several values of the fields $h$. Firstly, to approximate the limiting
value $v_0 \approx \infty $ or $\xi_{w\beta } \approx 0$, $v_0$ is set at 
$v_0=100$. 
In Figure $1$, $\langle y \rangle$ is plotted against $\ln (1/h)$ as in 
eq.(\ref{meanl}). The values 
of $\langle y \rangle$  lie on a straight-line and a logarithmic regression
gives a slope $\theta = 1.43 \pm 0.008$. 
The other limit,  $v_0 \approx 0 $ or $\xi_{w\beta } \approx \infty $, is 
approximated by $v_0 =0.05$. For smaller values of $v_0 $ the fluctuations
become very large, the autocorrelations increase and the simulations are not 
reliable. The values of $\langle y \rangle$, plotted 
against $\ln (1/h)$ in Figure $1$, show that they lie 
on a straight-line, but now with a slope $\theta =1.65 \pm 0.02$.
Figure $1$ suggest that $\theta $ is $v_0$ dependent and other 
intermediate values of $v_0$ are chosen to show this.
In Figure $2$ are plotted the fitted values of the slopes $\theta $ 
corresponding to different values of $v_0$ for lattice sizes $L=20, 40$. 
Importantly, the graph shows that the value of $\theta $ is indeed larger than 
the CW prediction $\theta^{CW}\approx 1.4$ for finite values of $v_0$, consistent
with (\ref{theta}). Due to the dependence on the 
cutoff $\Lambda $ a precise quantitative comparison between simulation and 
theory is not possible. Nevertheless a fit of the numerical results to the 
theoretical prediction (\ref{theta}) yields a value for the cutoff $\Lambda $
close to unity, which is encouraging. We also note that as $v_0 $ increases,
the increment $\Delta \theta $ decreases and we recover the CW model 
prediction $\theta^{CW} \approx 1.4 $ as $v_0 \rightarrow \infty $. 
For small $v_0 $ the measured value of $\theta $ is considerably bigger 
than the CW result and increases non-linearly with $1/v_0$. 
As mentioned above, the limiting value should be $\theta =1.9$, but this 
can not be reached in the simulations since it corresponds to infinite 
fluctuations.
Unfortunately, extrapolation to $v_0 =0$ is not possible either, given the 
non-linear dependence of $\theta $ on $v_0$ and large finite-size effects 
which would occur in this limit. Nevertheless the 
observed increase of $\theta $ with $v_0 $ testifies to the strong influence 
of the coupling between the two fields and is certainly consistent with 
the basic prediction of the coupled Hamiltonian theory.\\
In the light of [PB] theory \cite{PB}, the results are 
interpreted as follow: for very large values of $v_0$ the lower surface 
is very stiff and it does not fluctuate, hence the capillary wave result of 
$\theta^{CW} \approx 1.4$ is recovered. For decreasing values of 
$v_0$, even small fluctuations in the lower surface coupled to the 
upper surface produce a change in the critical amplitude  $\theta $ as 
predicted by the theory and also in agreement with the Ising model simulations.
This is the first time that RG predictions of an effective Hamiltonian, 
simulations on the Ising model and simulations of the same effective 
Hamiltonian are in qualitatively and quantitatively agreement. 
It would be interesting to extend this work to critical wetting and 
investigate if (\ref{twoham}) can shed light on the controversy there. 
However this is beyond the scope of the present letter.
 
\section*{Acknowledgements}

I would like to thank Andrew Parry for suggesting this work, for 
fruitful discussions and for a critical reading of the manuscript.

\newpage

\begin{figure}[h]
\begin{center}
\scalebox{0.7}{\rotatebox{-90}{\includegraphics{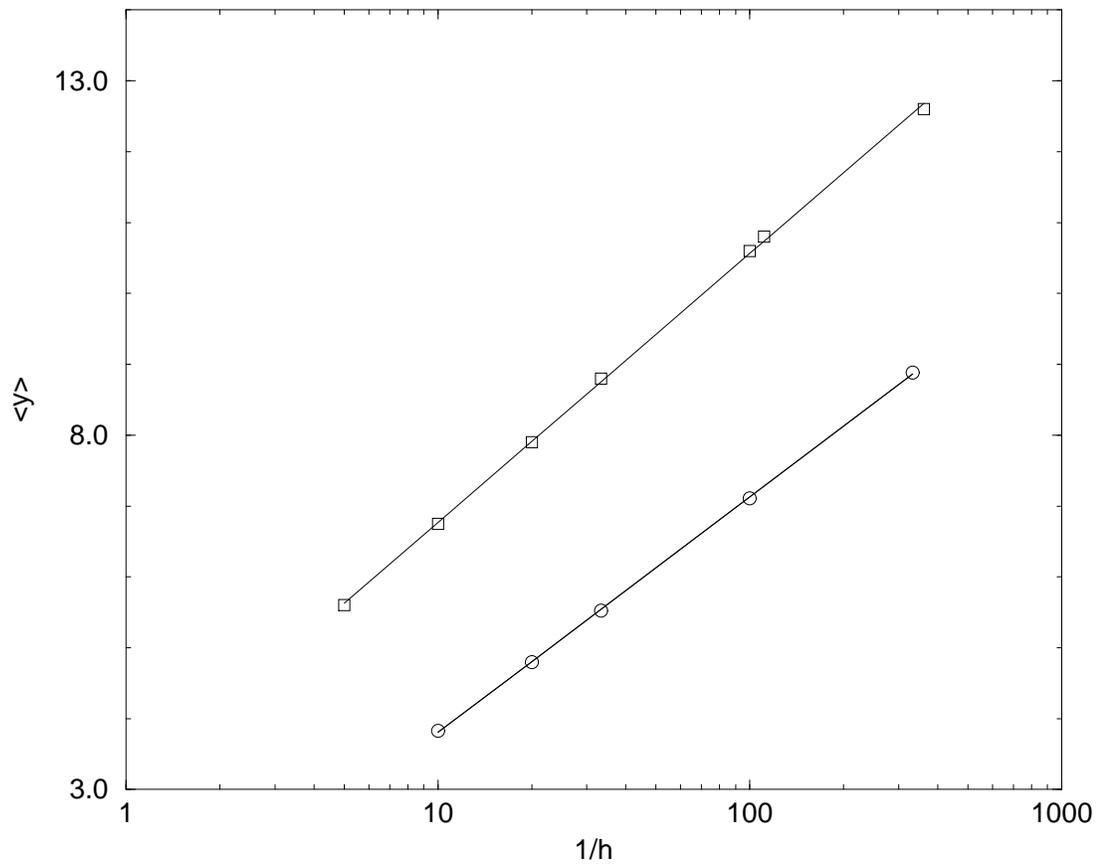}}}
\end{center}
\caption{Mean interface displacement $\langle y \rangle$ against $\log (1/h)$,
for $v_0=100$ (circles) and $v_0=0.05$ (squares).
The error bars are within the symbols.
The solid lines are the logarithmic regressions.}
\end{figure}

\begin{figure}[h]
\begin{center}
\scalebox{0.7}{\rotatebox{-90}{\includegraphics{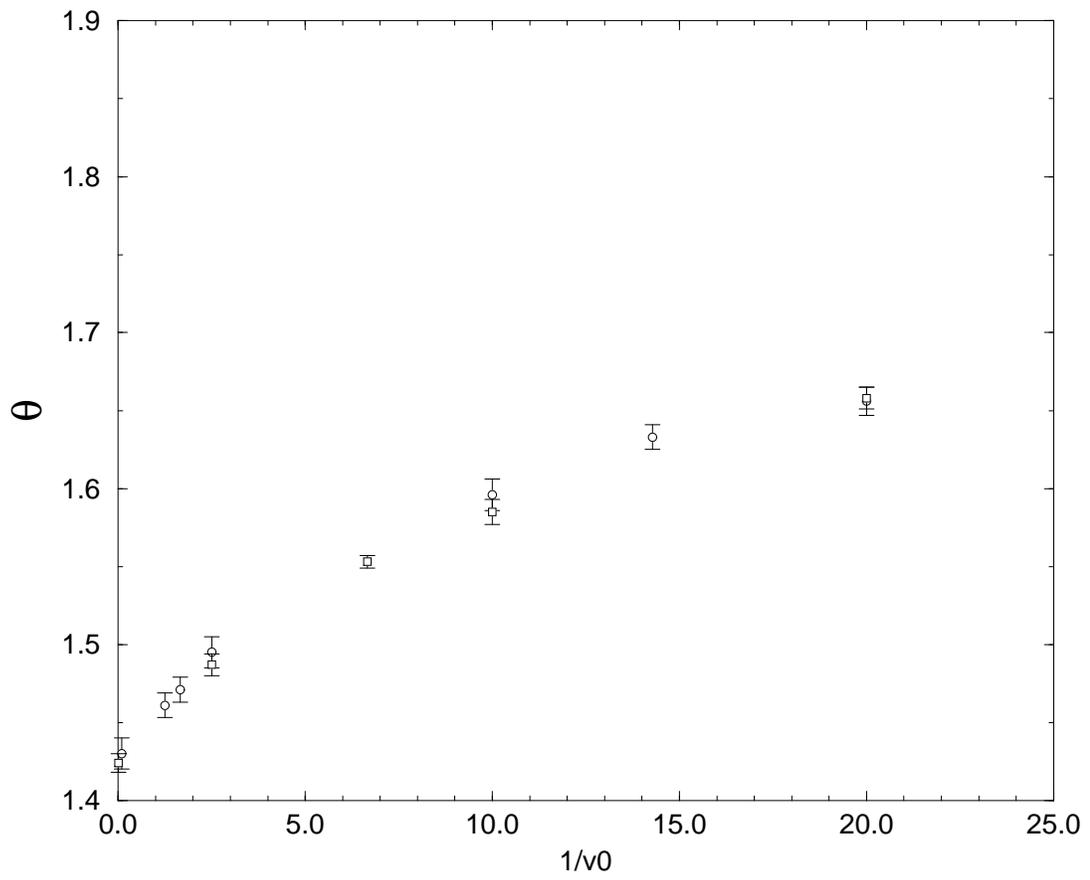}}}
\end{center}
\caption{Critical amplitude $\theta $ as a function of the inverse 
parameter $v_0$, for lattice size $L=20$ (circles) and $L=40$ (squares).}
\end{figure}

\end{document}